\documentclass[12pt]{article}
\usepackage{epsf,amsfonts,amssymb,amsbsy}
\usepackage{graphicx}
\def\pacs{\noindent PACS:~}
\begin{document}
\title{Spinor algebra transformations as gauge symmetry:\\
limit to Einstein gravity}
\date{}
\author{V.V.Kiselev,\\~ \normalsize
State Research Center "Institute for High Energy Physics" \\
{~ \normalsize Protvino, Moscow region, 142280 Russia}\\
~ \normalsize Fax: +7-095-2302337, E-mail: kiselev@th1.ihep.su
}

%
\maketitle

\begin{abstract}
We construct a Lagrangian of Weyl spinors and gauge fields, which is invariant
under the action of equivalent local transformations on the spinor algebra
representations. A model of vacuum with a nontrivial gauge strength-tensor
setting a scale and spontaneously breaking the gauge symmetry is suggested, so
that the leading approximation at low energies is described by the
Einstein--Hilbert Lagrangian of gravity. We consider a mechanism for the
cancellation of cosmological constant due to a symmetry between the vacuum
strength tensor and the tensor dual to it. The appearance of nonzero masses for
the non-graviton degrees of freedom for the gauge field is shown. The
generalization to the Dirac spinors is considered, and a group of additional
gauge symmetry is determined. We argue for a necessary introduction of
supersymmetry.
\end{abstract}


\section{Introduction}
For two-component Weyl spinors $\theta_\alpha$ the algebra of 
$\sigma$-matrices is defined by the commutation relations 
\begin{equation}
\begin{array}{rcl}
\{\sigma^\mu\bar\sigma^\nu + \sigma^\nu\bar\sigma^\mu\}_\alpha^{~\beta} &=& 2
\eta^{\mu\nu} \, \delta_\alpha^\beta, \\[4mm]
\{\bar \sigma^\mu \sigma^\nu + \bar
\sigma^\nu\sigma^\mu\}^{\dot\alpha}_{~\dot\beta} &=& 2
\eta^{\mu\nu} \, \delta^{\dot\alpha}_{\dot\beta},
\end{array}
\label{algebra}
\end{equation}
where $\eta^{\mu\nu}={\rm diag}[1,-{\bf 1}]$ is the metric tensor of
Minkowski, and the Hermitian matrices $\sigma^{\mu} = (1,{\boldsymbol \sigma})$
have the explicit indices: $\sigma^{\mu}_{\alpha\dot\alpha}$ and
$\bar\sigma^{\mu\dot\alpha\alpha}= \sigma^{\mu}_{\beta\dot\beta}\,
\epsilon^{\dot\alpha\dot\beta}\, \epsilon^{\alpha\beta}$, so that $\epsilon$ is
completely anti-symmetric tensor with the normalization $\epsilon^{12}=1$, and
the spinor indices are defined in accordance with the following prescriptions
accepted in \cite{BW}:

\begin{description}
\item[~~]
$\theta_\alpha$ is the two-component left-handed spinor-column,
\item[~~]
$\bar\theta_{\dot\alpha}= [\theta_\alpha]^\dagger$ is the Hermitian-conjugated
spinor-row,
\item[~~]
$\bar\theta^{\dot\alpha}= \epsilon^{\dot\alpha\dot\beta} [\theta_\beta]^\star
=\theta_c$ is the charge-conjugated spinor-column,
\item[~~]
$\theta^{\alpha} = \epsilon^{\alpha\beta} \theta_\beta = [\theta_c]^\dagger$
is the Hermitian-conjugated spinor-row of charge-conjugated spinor.
\end{description}
Then the invertible complex matrices with the unit determinant $M \in SL(2,
{\boldsymbol C})$ transform the spinors in the following way:

\begin{description}
\item[~~]
$\theta^\prime_\alpha = M_{\alpha}^{~\beta}\, \theta_{\beta}$, or in the matrix
notations $\theta^\prime = M \theta$,
\item[~~]
$\bar\theta^\prime_{\dot\alpha} = M_{~\dot\alpha}^{\star~\dot\beta}\,
\bar\theta_{\dot\beta}$, i.e. $\bar\theta^\prime = \bar \theta M^\dagger$,
\item[~~]
$\bar\theta^{\prime\dot\alpha} = [M^\star]_{~~\dot\beta}^{-1~\dot\alpha}\,
\bar\theta^{\dot\beta}$, i.e. $\theta_c^\prime = [M^\dagger]^{-1} \theta_c$,
\item[~~]
$\theta^{\prime\alpha} = [M^{-1}]_{\beta}^{~~\alpha}\,
\theta^{\beta}$, i.e. $[\theta_c^\prime]^\dagger = [\theta_c]^\dagger  M^{-1}$,
\end{description}
so that we can easily show that the products of $\theta_c^\dagger \theta =
\theta^\alpha \theta_\alpha$ and $\theta^\dagger \theta_c = \bar\theta_{\dot
\alpha}\bar\theta^{\dot\alpha}$ are invariant. The matrices $M$ generate the
equivalent transformations of $\sigma$-matrix representations, conserving the
form of commutation relations (\ref{algebra}):
\begin{equation}
\sigma^{\prime\mu} = M\, \sigma^\mu \, M^\dagger,\;\;\;\; 
\bar\sigma^{\prime\mu} = [M^\dagger]^{-1}\, \bar\sigma^\mu \, M^{-1}.
\end{equation}
The infinitesimal transformations are given by six components of anti-symmetric
tensor $\omega_{nm}$, where $\{n,m\} = (0,1,2,3)$, so that
\begin{equation}
\begin{array}{rcl}
M &=& 1 + \sigma^{nm} \omega_{nm},\\[2mm]
[M^\dagger]^{-1} &=& 1 + \bar\sigma^{nm} \omega_{nm},
\end{array}
\;\;\;\;\; \omega_{nm}\to 0,
\label{inf}
\end{equation}
and the generators of transformations are defined by the following relations:
\begin{equation}
\begin{array}{rcl}
[\sigma^{nm}]_{\alpha}^{~\beta} &=& \frac{1}{4}\left[
\sigma^{n}_{\alpha\dot\alpha}\bar\sigma^{m\dot\alpha\beta} - 
\sigma^{m}_{\alpha\dot\alpha}\bar\sigma^{n\dot\alpha\beta}\right],\\[2mm]
[\bar\sigma^{nm}]^{\dot\alpha}_{~\dot\beta} &=& \frac{1}{4}\left[
\bar\sigma^{n\dot\alpha\beta}\sigma^{m}_{\beta\dot\beta} - 
\bar\sigma^{m\dot\alpha\beta}\sigma^{n}_{\beta\dot\beta}\right].
\end{array}
\end{equation}
In contrast to the space-time transformations by the Lorentz group, which are
given by the tensor of busts and rotations $\omega$ and act on the Weyl spinors
by the corresponding matrices of $M(\omega)$ in the theory of special
relativity, let us consider the equivalent transformations of spinor algebra
representations as a gauge group, which action does not change the space-time
coordinates. Such the gauge group changes the explicit form of
$\sigma$-matrices, which can differ from the standard definition of Pauli
matrices. The observables of free Weyl particle do not depend on the operation
of this group, of course, i.e. the group transformations conserve the invariant
action of Weyl spinor:
\begin{equation}
\begin{array}{rcl}
S &=& \int d^4x\; {\cal L}_0,\\[4mm]
{\cal L}_0 &=& \frac{i}{2} [\bar \theta(x)\, \bar \sigma^{\mu}\partial_\mu
\theta(x) + 
\theta(x)\, \sigma^{\mu}\partial_\mu \bar \theta(x) ].
\end{array}
\end{equation}
Thus, we study the principle of relativity for the choice of basis system
in the algebra of $\sigma$-matrices for the Weyl spinors.

In the present paper we consider the global transformations of algebra
representations for the Weyl spinors in the Minkowskian space-time and derive
the corresponding Noether currents in section 2. These currents determine the
form of interaction with external sources, so that this interaction has the
form of product for the current and spin connection, which appears in the
description of gravitational interaction for the particles possessing the spin
in addition to the contact term of energy-momentum tensor with the metrics.
Further we study the local gauge invariance in the spinor algebra and introduce
the covariant derivative on the left-handed and right-handed Weyl spinors in
section 3. The commutator of covariant derivatives determines the tensor of
gauge field strength, which coincides with the curvature tensor of
spin-connection. We formulate the action of gauge connection-field in the
Minkowskian space-time in terms of standard Lagrangian written down as the
square of strength tensor. This Lagrangian allows the expression in the form of
trace for the square of gauge field strength tensor in the algebra of
generators on the left-handed and right-handed spinors. We assume that the
Lagrangian of spin-connection gauge field interacting with the Weyl spinors is
renormalizable\footnote{The renormalizability of gravity with the terms
quadratic in the curvature tensor was demonstrated by K.S.Stelle in
\cite{stelle}.} in the Minkowskian space-time, as it usually takes place in the
gauge theories. The invariance of Lagrangian under the action of group is
provided by introducing the gauge transformation of tetrad (vierbein), which
determine the relation between the basis in the spinor algebra and the system
of Minkowskian space-time coordinates. As a consequence of tetrad
transformation, the energy-momentum tensor by its anti-symmetric part enters
the law for the conservation of spinor current. The local symmetry leads to the
introduction of {\it auxiliary} field of tetrad\footnote{In this way, the
corresponding tensor of curvature under the metric connection generally can
remain equal to zero, and the space-time of Minkowski is not curved as referred
to the spinor algebra.}. However, we have two kinds of coordinate indices
related to the ``world'' points and local Minkowskian reference-system
connected to the algebra of $\sigma$-matrices. Therefore, the invariant measure
of space-time volume defined under the ``world'' coordinates should contain the
usual factor of ${\rm det} [h]$, where $h$ denotes the tetrad, which can
determine a Riemannian metric.

The idea explored in the present paper is connected to that of obtaining the
general relativity by gauging the Lorentz group as date back to Utiyama
\cite{Utiyama}, and that approach has been widely used in supergravity, where
it is the simplest way to find the appropriate invariants. The essential
difference of present consideration from that of Utiyama is that we do not deal
with the transformation of coordinates, while we investigate the local
symmetry, which does not act on the space-time variables. A similar model of
extended Yang--Mills gauge theory in Euclidean space was considered in
\cite{Gabrielli}, where the author focused on the description of higher spins
with renormalizable interactions including the supersymmetry and spontaneous
breaking of local symmetry in the Higgs mechanism with scalar particles, which
is rather different from the approach explored in the persent paper. So, the
material upto section \ref{3.1} gives a presentation of rather known things as
suitable for the purposes of current study.

Further we involve an assumption on a gauge field condensation and the
spontaneous breaking of gauge symmtery, that results in a background strength
tensor in the classic theory. Particularly, we write down an example of global,
independent of coordinates spin-connection with a covariant expectation value
of vacuum field, that can be related with the expansion of strength tensor in
terms of sum for the strength tensor of vacuum background field and the
dynamical tensor of curvature. The contribution linear over the background
strength and dynamical curvature coincides with the form of Lagrangian in the
Einstein--Hilbert theory of gravitation. Thus, we see that the action of
general relativity occurs as the result of the gauge field condensation and the
spontaneous breaking of gauge symmtery, and in this limit it is the effective
low-energy contribution into the full action, that leads to the curved
space-time because of the vacuum fields. Note, that the condensation conserves
the local gauge symmetry and, hence, the local Lorentz invariance takes place
as well as the symmetry under the general coordinate transformations does, so
that the invariant measure in terms of ``world'' indices naturally contains the
ordinary factor of ${\rm det}[h]$. The term of Lagrangian quadratic over the
dynamical strength tensor becomes significant at the energy of spin-connection
quanta about the Planck scale in the vicinity of which, probably, the
decondensation takes place, and therefore the space-time becomes flat under the
absence of vacuum fields. The Lagrangian term quadratic over the background
connection strength tensor generally leads to a cosmological constant, which is
cancelled by the contribution caused by the product of background connection
and spinor current, if we assume that there is a corresponding condensate of
spinor field. The complete cancellation of these two terms can be broken by a
fluctuation of fields, that naturally leads to an inflation expansion or
contraction of universe depending on the sign of fluctuation. We show that in
the studied example of background connection, the dynamical modes of
excitations have the effective masses in the region of Planck scale. So, we
assume that the modes of gauge field except the graviton aer suppressed outside
the Planck scale in the analogous way to the gluons in QCD, and these modes can
be confined. Further we consider some problems appearing in the formulation of
Hamiltonian dynamics of spinor-connection gauge field and suggest an ansatz in
terms of dual fields, that allows us to classify the dynamical modes of field
and to offer an alternative mechanism for the cancellation of cosmological
constant due to a symmetry between the vacuum gauge field and the field in the
dual strength tensor. We comment also the problem on how the vacuum fields
enter the field equations.

In section 4 we briefly discuss modifications of approach under consideration
in the case of Dirac field, for which we naturally introduce an additional
gauge symmetry including the electroweak symmetry of standard model. In section
5 we discuss a reason for the correlation of vacuum bosonic and fermionic
fields in the light of supersymmetry, that, probably, provides the cancellation
of cosmological constant. We emphasize that the introduction of supersymmetry
is a necessary consequence of BRST-generalization of evolution operator for the
spinor field. Finally, we discuss the obtained results of suggested approach in
conclusion.

\section{Global symmetry. Noether currents}
Following the standard procedure, let us determine the Noether currents
according to the formula
\begin{equation}
j^{\mu}_a = \frac{\delta {\cal L}}{\delta \partial_\mu \psi}\;
\frac{\partial \psi}{\partial \omega^a} + 
\frac{\delta {\cal L}}{\delta \partial_\mu \bar \psi}\; \frac{\partial \bar
\psi}{\partial \omega^a},
\end{equation}
where in the case under study we put $\omega^a\to \omega_{nm}$, the parameters
of infinitesimal transformations (\ref{inf}) for the spinors $\theta$ and
$\bar\theta$. Then in the explicit form we find
\begin{equation}
j^{\mu,nm} = \frac{i}{2} [\bar \theta(x)\, \bar \sigma^{\mu}\, \sigma^{nm}
\theta(x) + 
\theta(x)\, \sigma^{\mu}\, \bar \sigma^{nm}\bar \theta(x) ].
\label{current}
\end{equation}
Using the Pauli gauge for the $\sigma$-matrices, $\sigma^{\mu} =
(1,{\boldsymbol \sigma})$, we can show that 
\begin{equation}
j^{\mu,nm} = \frac{1}{2}\, \epsilon^{\mu nm \nu}\, j^{\lambda} \,
\eta_{\nu\lambda},
\label{currents}
\end{equation}
where $ j^{\lambda} = \bar \theta\, \bar\sigma^{\lambda}\,\theta$ is the spinor
current, which is rotated under the action of group transformations in the
space of $\sigma$-matrices.

Under the action of global transformation we get the spinor Lagrangian equal to
\begin{equation}
\begin{array}{rcl}
{\cal L}_0^{\prime} &=& \frac{i}{2} [(\bar \theta M^\dagger)\,
([M^\dagger]^{-1}\bar \sigma^{\mu}M^{-1})\, \partial_\mu (M\theta) +
(\theta M^{-1})\, (M\sigma^{\mu}M^\dagger)\,\partial_\mu ([M^\dagger]^{-1}\bar
\theta) ] \\[3mm]
 &=& \frac{i}{2} [\bar \theta^\prime \, \bar \sigma^{\prime\mu}\partial_\mu
\theta^\prime + 
\theta^\prime\, \sigma^{\prime\mu}\partial_\mu \bar \theta^\prime ] = {\cal
L}_0,
\end{array}
\end{equation}
and it is quite evident that the variation of Lagrangian is equal to zero,
${\cal L}_0^\prime - {\cal L}_0=0$. Let us introduce the tetrad $h_m^{~\mu}$,
which relates the local ``latin'' indices in the basis of $\sigma$-matrices
with the world ``greek'' indices of space-time:
\begin{equation}
\begin{array}{lll}
\sigma^\mu = \sigma^m \, h_m^{~~\mu}, &
\bar\sigma^\mu = \bar\sigma^m \, h_{m}^{~~\mu}, &
h_m^{~~\mu}\, h_{n}^{~~\nu}\, \eta^{mn} = \eta^{\mu\nu},\\[3mm]
\sigma^m = \sigma^\mu \, h^{~m}_{\mu}, &
\bar\sigma^m = \bar\sigma^\mu \, h^{~m}_{\mu}, &
h_{~~m}^{\mu}\, h_{~~n}^{\nu}\, \eta_{\mu\nu} = \eta_{mn},
\end{array}
\;\;\;\;h_m^{~~\mu}\, h^{\nu}_{~~n}\, \eta_{\mu\nu} = \eta_{mn},
\label{tetrada}
\end{equation}
so that for the transformed $\sigma$-matrices we get
\begin{equation}
\sigma^{\prime\mu} = \sigma^m \, h_m^{\prime~\mu},
\end{equation}
and the infinitesimal rotation of tetrad has the form
\begin{equation}
\delta h_m^{~\mu} = h_n^{~\mu}\, \omega_{lm}\, \eta^{nl}.
\end{equation}
The variation of inverse tetrad $h_{~n}^{\nu}$ can be written down in
accordance with (\ref{tetrada}) 
\begin{equation}
\delta h_{~n}^{\nu} = - h^{\nu m} h_{\mu n}\, \delta h_m^{~\mu},
\end{equation}
where the tetrad indices are moved down and up by the metric tensor and that
of inverse to it, correspondingly.

Then the variation of Lagrangian can be written down in the form
\begin{equation}
\delta{\cal L} = \delta\omega_{nm}\; [\partial_{\mu}\, j^{\mu,nm}+
h^{n\mu} T^{~m}_{\mu}],
\end{equation}
so that the invariance of Lagrangian under the global transformations leads to
the conservation law
\begin{equation}
\partial_{\mu}\, j^{\mu,mn}+\frac{1}{2} (T^{mn} - T^{nm})=0.
\label{conserv}
\end{equation}
As we see, in the absence of sources the current is conserved with the accuracy
up to the anti-symmetric part of energy-momentum tensor $T^{nm}$ for the spinor
field. This tensor could be generally reduced to the symmetric one due to the
additional term of Lagrangian in the form of divergence for a current,
so that this term is transformed to the surface integral and the equations of
motion remain with no changes.

After the introduction of sources the interaction Lagrangian has the
form\footnote{We see that the tetrad is not a source for a Noether current,
and therefore, it is an auxiliary field.}
\begin{equation}
{\cal L}_{\rm int} = j^{\mu,nm}\, {\cal A}_{\mu,nm}.
\end{equation}
In this way the source, i.e. the spin-connection, should possess some
transformation properties under the action of symmetry group for the full
action remains invariant after the change of spinor algebra representation.
These properties of source are investigated in the study of local symmetry.

\section{Local symmetry}

The generators of spinor transformations satisfy the standard commutation
relations
\begin{equation}
\begin{array}{rcl}
[\sigma^{nm},\, \sigma^{kl}] &=& - \eta^{nk}\sigma^{ml} + \eta^{nl}\sigma^{mk}
- \eta^{ml}\sigma^{nk} + \eta^{mk}\sigma^{nl}, \\[3mm]
[\bar\sigma^{nm},\, \bar\sigma^{kl}] &=& - \eta^{nk}\bar\sigma^{ml} +
\eta^{nl}\bar\sigma^{mk} - \eta^{ml}\bar\sigma^{nk} +
\eta^{mk}\bar\sigma^{nl},
\end{array}
\end{equation}
so that we introduce the ``covariant'' derivatives on the left-handed and
right-handed spinor fields
\begin{equation}
\begin{array}{rcl}
\nabla_{\mu\alpha}^{~~~\beta} &=& \delta_\alpha^{~\beta}\, \partial_\mu + 
{\cal A}_{\mu,nm} \sigma^{nm~\beta}_{~~\alpha}, \\[3mm]
\overline\nabla_{\mu~\dot\beta}^{~\dot\alpha} &=&
\delta^{\dot\alpha}_{~\dot\beta}\, \partial_\mu +  {\cal A}_{\mu,nm}
\bar\sigma^{nm\dot\alpha}_{~~~~~\dot\beta},
\end{array}
\end{equation}
which have the consistent commutator determining the strength tensor of gauge
field ${\cal A}$
\begin{equation}
\begin{array}{rcl}
[\nabla_\mu,\, \nabla_\nu] &=& {\cal F}_{\mu\nu, mn}\, \sigma^{mn},\\[3mm]
[\overline\nabla_\mu,\, \overline\nabla_\nu] &=& {\cal F}_{\mu\nu, mn}\,
\bar\sigma^{mn},
\end{array}
\label{tensor}
\end{equation}
where
\begin{equation}
{\cal F}_{\mu\nu, mn} = \partial_\mu {\cal A}_{\nu,mn} - \partial_\nu {\cal
A}_{\mu,mn} + 2 ({\cal A}_{\mu,mk} {\cal A}_{\nu,ln} - {\cal A}_{\nu,mk} {\cal
A}_{\mu,ln})\, \eta^{kl}.
\label{strength}
\end{equation}
In the derivation of (\ref{strength}) we have explored the anti-symmetry of
spin-connection over the group indices ${\cal A}_{\mu,nm} = - {\cal
A}_{\mu,mn}$. We emphasize that the strength tensor is reduced to the standard
form of curvature tensor for the connection determined by $\Gamma_{\mu,nm} =
{2}{\cal A}_{\mu,nm}$, so that ${\cal F}_{\mu\nu, mn}[{\cal A}] = \frac{1}{2}
{\cal R}_{\mu\nu,nm}[\Gamma]$.

Define the Lagrangian of gauge field ${\cal A}$ in the space-time of Minkowski
in the ordinary form\footnote{The sign in front of strength tensor squared is
fixed in the consistent way along with the definition of covariant derivative.} 
\begin{equation}
{\cal L}_{\cal A} = \frac{1}{4 g^2_{\rm\scriptscriptstyle Pl}}\, {\cal
F}_{\mu\nu, mn} {\cal F}^{\mu\nu, mn},
\label{l-action}
\end{equation}
where $g_{\rm\scriptscriptstyle Pl}$ is a coupling constant of gauge field.
Let us show that Lagrangian (\ref{l-action}) is invariant under the action of
local gauge transformations  giving the equivalent representations of spinor
algebra, if we introduce the following transformations of gauge fields:
\begin{equation}
{\cal A}^{\prime}_{\mu,nm}\, \sigma^{nm} = M^{-1} {\cal A}_{\mu,nm}\,
\sigma^{nm}\, M - (\partial_\mu M^{-1})\cdot M,
\label{gauge}
\end{equation}
which follows from the definition of covariant derivative
\begin{equation}
\begin{array}{rcl}
M\, \nabla_\mu({\cal A}^\prime)\, M^{-1} &\stackrel{\rm\scriptscriptstyle
def}{=}& \nabla_\mu({\cal A}),\\[3mm]
[M^\dagger]^{-1}\, \overline\nabla_\mu({\cal A}^\prime)\, M^\dagger
&\stackrel{\rm\scriptscriptstyle def}{=}& \overline\nabla_\mu({\cal A}).
\end{array}
\label{def}
\end{equation}
Definitions (\ref{def}) are consistent, i.e. they result in the same
transformation of field ${\cal A}$ (\ref{gauge}), since the relation 
$[\sigma^{nm}]^\dagger = - \bar\sigma^{nm}$ is valid. Then, definition
(\ref{tensor}) leads to the group transformation of strength tensor in the
algebra of left-handed and right-handed spinors in the following form:
\begin{equation}
\begin{array}{rcl}
M\, {\cal F}_{\mu\nu, mn}({\cal A}^\prime)\, \sigma^{mn}\, M^{-1} &=&  {\cal
F}_{\mu\nu, mn}({\cal A})\, \sigma^{mn} = {\cal F}_{\mu\nu},\\[3mm]
[M^\dagger]^{-1}\, {\cal F}_{\mu\nu, mn}({\cal A}^\prime)\, \bar\sigma^{mn}\,
M^{\dagger} &=&  {\cal F}_{\mu\nu, mn}({\cal A})\, \bar\sigma^{mn} =
\overline{\cal F}_{\mu\nu}.
\end{array}
\end{equation}
Thus, the traces over the spinor indices for the squares of following
quantities are invariant:
\begin{equation}
\begin{array}{rcl}
{\rm Tr} [{\cal F}_{\mu\nu} {\cal F}^{\mu\nu}] &=& - \frac{1}{2}[\delta^m_k
\delta^n_l - \delta^m_l \delta^n_k - i \epsilon^{mn}_{~~~kl}]\, 
{\cal F}_{\mu\nu, mn} {\cal F}^{\mu\nu, kl}, \\[3mm]
{\rm Tr} [\overline{\cal F}_{\mu\nu} \overline{\cal F}^{\mu\nu}] &=& -
\frac{1}{2}[\delta^m_k \delta^n_l - \delta^m_l \delta^n_k + i
\epsilon^{mn}_{~~~kl}]\, {\cal F}_{\mu\nu, mn} {\cal F}^{\mu\nu, kl}.
\end{array}
\end{equation}
Then we can write down the Hermitian-conjugated expression 
\begin{equation}
\begin{array}{rcl}
{\cal L}_{\cal F} &=& - \Re {\mathfrak e}\, \frac{\displaystyle
1}{\displaystyle 8{\mathfrak g}_{\rm\scriptscriptstyle Pl}^2}\, {\rm Tr} [{\cal
F}_{\mu\nu} {\cal F}^{\mu\nu}] - \Re {\mathfrak e}\, \frac{\displaystyle
1}{\displaystyle 8{\mathfrak g}_{\rm\scriptscriptstyle Pl}^{\star 2}}\, {\rm
Tr} [\overline{\cal F}_{\mu\nu} \overline{\cal F}^{\mu\nu}] \\[5mm] 
&=& \frac{\displaystyle 1}{\displaystyle 4 g^2_{\rm\scriptscriptstyle
Pl}}\, {\cal F}_{\mu\nu, mn} {\cal F}^{\mu\nu, mn} + \frac{\displaystyle
\theta}{\displaystyle 32\pi^2}
\epsilon^{mn}_{~~~kl}\, {\cal F}_{\mu\nu, mn} {\cal F}^{\mu\nu, kl},
\end{array}
\label{LF}
\end{equation}
where we have introduced the notations for the real and imaginary parts of
inverse charge ${\mathfrak g}_{\rm\scriptscriptstyle Pl}$ squared, so that
\begin{equation}
\Re {\mathfrak e}\frac{1}{4 {\mathfrak g}_{\rm\scriptscriptstyle Pl}^2} =
\frac{1}{4 {g}_{\rm\scriptscriptstyle Pl}^2}, \;\;\;\; 
\Im {\mathfrak m}\frac{1}{4 {\mathfrak g}_{\rm\scriptscriptstyle Pl}^2} =
\frac{\theta}{16\pi^2}.
\end{equation}
In Lagrangian (\ref{LF}) we will not consider the ``$\theta$-term'', which is
usually introduced in the case of nontrivial structure of vacuum caused by
instantons\footnote{We do not concern for problems connected to the search for
nontrivial classical solutions in the Euclidean space. These solutions imply
the instanton amplitudes for the transitions between the vacua in the quantum
theory in the Minkowskian space-time. In addition, in the case under
consideration the ``duality'' is written down for the group indices, while
there is a possibility for the introduction of terms in the form of
$\epsilon^{\mu\nu\lambda\gamma} {\rm Tr} [{\cal F}_{\mu\nu} {\cal
F}_{\lambda\gamma}]$. Another note is that the group $SL(2,{\bf C})$ contains
the subgroup $SU(2)$, so that the construction of classical solutions could be
done by means of simple generalizing the case of $SU(2)$, though in this way
one should investigate possibilities for a non-invariance of solutions, since
additional transformations in the group of $SL(2,{\bf C})$ could lead to a
trivialization, i.e. a transformations of ``winding number''.}.

Then, the Lagrangian invariant under the local transformations of spinor
algebra representations has the form 
\begin{equation}
{\cal L} = \frac{1}{4 g^2_{\rm\scriptscriptstyle Pl}}\, {\cal
F}_{\mu\nu, mn} {\cal F}^{\mu\nu, mn} + \frac{i}{2} [\bar \theta(x)\, \bar
\sigma^{\mu}\nabla_\mu \theta(x) + 
\theta(x)\, \sigma^{\mu}\overline\nabla_\mu \bar \theta(x) ].
\label{ls-action}
\end{equation}

\subsection{Identities of Slavnov--Taylor}
\label{3.1}
The law of spinor current conservation in the presence of gauge field takes the
form
\begin{equation}
\partial_{\mu}\, j^{\mu,mn}+j^{\nu,kl}\, \left.\frac{\delta{\cal
A}_{\nu,kl}}{\delta
\omega_{mn}}\right|_{\rm\scriptscriptstyle glob} +\frac{1}{2}
(T^{mn} - T^{nm})=0,
\label{Aconserv}
\end{equation}
so that we introduce the covariant divergence of current by the following
definition:
\begin{equation}
[\nabla_\mu\, j^{\mu}]^{nm} = \partial_{\mu}\, j^{\mu,nm}+j^{\nu,kl}\,
\left.\frac{\delta{\cal A}_{\nu,kl}}{\delta \omega_{nm}}
\right|_{\rm\scriptscriptstyle glob}.
\end{equation}
The infinitesimal global transformations of gauge field can be written down in
the explicit form
\begin{equation}
\left.\frac{\delta{\cal A}_{\nu,kl}}{\delta \omega_{mn}}
\right|_{\rm\scriptscriptstyle glob} =  ({\cal A}_{\nu,kp} S^{mn}_{ql} - 
S^{mn}_{kp} {\cal A}_{\nu,ql})\,\eta^{pq},
\end{equation}
where the spin operator of vector particle is equal to
\begin{equation}
S^{mn}_{kl} = \delta^m_k \delta^n_l - \delta^m_l \delta^n_k.
\label{unit}
\end{equation}
The infinitesimal transformation of gauge field can be written down in the form
of covariant derivative for the transformation parameter $\omega$:
\begin{equation}
\delta_\omega {\cal A}_{\mu,kl} = [\nabla_\mu \omega]_{kl} = \partial_\mu
\omega_{kl} +( {\cal A}_{\mu,kp} S^{mn}_{ql} - S^{mn}_{kp} {\cal
A}_{\mu,ql})\,\eta^{pq}\, \omega_{mn}.
\label{nabla}
\end{equation}
So, we arrive to the following form of conservation law for the spinor current:
\begin{equation}
[\nabla_\mu\, j^{\mu}]^{mn} +\frac{1}{2} (T^{mn} - T^{nm})=0.
\label{Nconserv}
\end{equation}
Introduce external sources ${\cal J}$ for the gauge field ${\cal A}$:
$$
{\cal L}_{\cal J} = {\cal A}_{\mu,mn} {\cal J}^{\mu,mn}.
$$
Let us fix the gauge in the Lorentz form, for example:
\begin{equation}
\partial^{\mu} {\cal A}_{\mu,mn} = 0,
\label{gf}
\end{equation}
and, following the ordinary procedure, add the gauge fixing term to the
Lagrangian
\begin{equation}
{\cal L}_{\rm\scriptscriptstyle gf} = \frac{1}{2\alpha}\, (\partial^\mu{\cal
A}_{\mu,mn})^2.
\end{equation}
The full Lagrangian for the gauge fields ${\cal L} = {\cal L}_{\cal A} +{\cal
L}_{\rm\scriptscriptstyle gf} + {\cal L}_{\cal J}$ remains invariant, if for
any parameters of transformations $\omega$ we have
\begin{equation}
\left\{\frac{\delta ({\cal L}_{\rm\scriptscriptstyle gf} + {\cal L}_{\cal
J})}{\delta \omega}\, \omega = 0,\;\;\; \forall \omega \right\} \Rightarrow
\;\;\; \frac{\delta ({\cal L}_{\rm\scriptscriptstyle gf} + {\cal L}_{\cal
J})}{\delta \omega} = 0.
\label{cond}
\end{equation}
Let us calculate the operator in (\ref{cond})
\begin{equation}
\frac{\delta ({\cal L}_{\rm\scriptscriptstyle gf} + {\cal L}_{\cal
J})}{\delta \omega}\, \omega= \frac{1}{\alpha}\, (\partial^{\mu}{\cal
A}_{\mu})\;
\partial^\nu [\nabla_\nu \omega]+ {\cal J}^{\mu} [\nabla_\mu \omega].
\label{oper}
\end{equation}
Introduce the operator inverse to ${\mathfrak M} ({\cal A})= \partial^\nu
\nabla_\nu({\cal A})$, and write down eq. (\ref{cond}) at $\omega =
{\mathfrak M}^{-1}({\cal A})\, \tilde \omega$, $\forall \tilde\omega$:
\begin{equation}
\frac{1}{\alpha}\, (\partial^{\mu}{\cal A}_{\mu}) + {\cal J}^{\mu}
[\nabla_\mu({\cal A})\, {\mathfrak M}^{-1} ({\cal A})]=0.
\label{ST}
\end{equation}
Eq. (\ref{ST}) gives the identities of Slavnov--Taylor or the generalized
identities of Ward--Takahashi. These equations are important in the quantum
theory for the prove of renormalization \cite{slavfad}. They are usually
written down for the partition functional dependent of sources $G[{\cal J}]$,
so that the field is replaced by the variational derivative ${\cal A} \equiv
\frac{\delta}{\delta{\cal J}}$:
\begin{equation}
\frac{1}{\alpha}\, \partial^{\mu}\frac{\delta G[{\cal J}]}{\delta{\cal
J}^{\mu}} + {\cal J}^{\mu} \left[\nabla_\mu\left(\frac{\delta}{\delta{\cal
J}}\right)\, {\mathfrak M}^{-1} \left(\frac{\delta}{\delta{\cal
J}}\right)\right] G[{\cal J}]=0.
\label{STquant}
\end{equation}
Thus, the identities of Slavnov--Taylor show, which constraints\footnote{In the
quantum theory these identities can be broken in the case of anomaly.} appear
for the Green functions due to the independence of observables on the gauge
degrees of freedom. In the quantum theory the partition functional is written
down in the form of continual integral
$$
e^{iG[{\cal J}]} = \int e^{i\int d^4x\, ({\cal L}_{\cal A} +{\cal
L}_{\rm\scriptscriptstyle gf} + {\cal L}_{\cal J})} {\rm det}[{\mathfrak M}]\,
{\cal D A},
$$
whereas one usually introduces the Grassmann ghosts by Faddeev--Popov
$$
{\rm det}[{\mathfrak M}] = \int e^{i\int d^4x\, {\cal L}_{\rm\scriptscriptstyle
gh}}\; {\cal D}c\,{\cal D}\bar c,
$$
where 
$$
{\cal L}_{\rm\scriptscriptstyle gh} = \bar c(x)\, \partial^\mu\nabla_\mu({\cal
A})\, c(x),
$$
so that the operator ${\mathfrak M}^{-1}({\cal A})$ is the propagator of ghosts 
$c_{mn}$ in the external field $\cal A$.

\subsection{Einstein gravity}
Assume that there is a potential in the effective action, which leads to
a nonzero vacuum strength of gauge fields. We suggest that the strength tensor
can be represented by a decomposition in terms of sum over the tensors of
some components, so that
\begin{equation}
{\cal F}_{\mu\nu, mn} = {\cal R}^{0[a]}_{\mu\nu, mn} + {\cal R}^{0[b]}_{\mu\nu,
mn} + \frac{1}{2}{\cal R}_{\mu\nu, mn}(\Gamma),
\label{vacF}
\end{equation}
so that the corresponding spin-connection components are determined by
expressions with the following kinds of Lorentz structures:
\begin{eqnarray}
{\cal A}_{\mu,mn}^{[a]} &=& \frac{1}{\sqrt{2}}\, \epsilon_{\mu\nu mn} a^\nu ,
\nonumber\\
{\cal A}_{\mu,mn}^{[b]} &=& \frac{1}{\sqrt{2}}\,
(\eta_{\mu m} b_n - \eta_{\mu n} b_m), \label{vac}\\
{\cal A}_{\mu,mn}^{[\Gamma]} &=&  \frac{1}{2}\, \Gamma_{\mu,mn},
\nonumber
\end{eqnarray}
where the vacuum fields $a$ and $b$ have the expectation values
\begin{equation}
\begin{array}{rcl}
\langle a^\mu\rangle = 0,\;\;\; \langle b^\mu\rangle = 0,\;\;\; 
&&\langle a^\mu a^\nu \rangle = \frac{1}{4}\, \eta^{\mu\nu}
g^2_{\rm\scriptscriptstyle Pl} v_a^2,\;\;\; \\[4mm] 
\langle a^\mu b^\nu \rangle = \frac{1}{4}\, \eta^{\mu\nu}
g^2_{\rm\scriptscriptstyle Pl} v_a \, v_b,\;\;\; &&
\langle b^\mu b^\nu \rangle = \frac{1}{4}\, \eta^{\mu\nu}
g^2_{\rm\scriptscriptstyle Pl} v_b^2 .
\end{array}
\end{equation}
Thus, we have introduced two parallel time-like vectors, so that the
corresponding connections are dual over the group indices: ${\cal
A}_{\mu,mn}^{[b]} = - \frac{1}{2} \epsilon_{mnkl}\, {\cal A}^{~~~kl}_{[a]\mu}$.

Let us emphasize that the prescriptions in (\ref{vac}) determine the components
of strength tensor, which is valid under a gauge condition. Of course, we can
apply the general gauge transformations to the components in (\ref{vac}), so
that due to the linear decomposition in (\ref{vacF}) the components of strength
tensor are transformed independently. This fact implies that we deal with the
presentation, where the gauge symmetry is not broken, while the strength tensor
of gauge fields acquires the nontrivial vacuum expectation, that implies the
spontaneous breaking of gauge symmtery.

Then, for the strength tensor of vacuum we have got
\begin{equation}
\begin{array}{rcl}
{\cal R}^{0[a]}_{\mu\nu, mn}(a) &=& - (\epsilon_{\mu\alpha
mk}\epsilon_{\nu\beta nk} - \epsilon_{\mu\alpha nk}\epsilon_{\nu\beta mk})
a^{\alpha} a^{\beta}\\[3mm]
 &=& a^2 \eta_{\mu m}\eta_{\nu n} - a^2 \eta_{\nu m}\eta_{\mu n} - a_\nu a_n
 \eta_{\mu m} + a_\nu a_m \eta_{\mu n} + a_\mu a_n \eta_{\nu m} - a_\mu a_m
 \eta_{\nu n},\\[3mm]
{\cal R}^{0[b]}_{\mu\nu, mn}(b) &=& - {\cal R}^{0[a]}_{\mu\nu, mn}(b) .
\end{array}
\end{equation}
The term of Lagrangian linear over the background strength tensor and dynamical
tensor of curvature is equal to
\begin{equation}
\begin{array}{rcl}
{\cal L}_G &=&  \frac{\displaystyle 1}{\displaystyle g_{\rm\scriptscriptstyle
Pl}^2}\, {\cal R}_{\mu \nu}\cdot \left\langle\left(\frac{1}{2} a^2
\eta^{\mu\nu} - a^\mu a^\nu\right)-\left(\frac{1}{2} b^2 \eta^{\mu\nu} - b^\mu
b^\nu\right)\right\rangle \\[5mm]
&=& \frac{\displaystyle 1}{\displaystyle g_{\rm\scriptscriptstyle Pl}^2}\,
\left( \frac{1}{2}{\cal R}\, \eta_{\mu\nu} - {\cal R}_{\mu \nu}\right)\,
\langle a^\mu a^\nu - b^\mu b^\nu \rangle = -\frac{1}{4}\, (v_b^2-v_a^2)\;
{\cal R}(\Gamma),
\end{array}
\label{gravity}
\end{equation}
where we have introduced the Ricci tensor ${\cal R}_{\mu\nu}(\Gamma) = {\cal
R}_{\gamma\mu mn}(\Gamma)\, h^{m\gamma}\, h^{~n}_\nu$ and the scalar curvature
${\cal R}(\Gamma) = {\cal R}_{\mu\nu}\, \eta^{\mu\nu}$. Sure, the Lagrangian of
(\ref{gravity}) strictly coincides with the Lagrangian of Einstein--Hilbert
gravity in the theory of general relativity, if we denote the gravitational
constant $\kappa = \frac{1}{v_b^2-v_a^2}$ and add the factor defining the
invariant measure ${\rm det}[h]\, d^4 x$ depending on the tetrad $h_{m\mu}$. 

Here we challenge the role of tetrad. We have introduced the tetrad as an
auxiliary field, which has been locally and globally reduced to the unit symbol
of Kronecker in the Minkowskian space-time. Under the gauge field condensation
implying the spontaneous breaking of gauge symmtery, the nontrivial vacuum
configuration leads to that the tetrad becomes a field depending on the
coordinates of space-time, so that it is only locally reduced to the unit under
the action of general coordinate transformations, while we explore the linear
approximation of (\ref{gravity}), and the dynamical characteristics of
connection $\Gamma$ can be ordinary reassigned to the characteristics of tetrad
or to the world metric tensor constructed by the tetrad.

Indeed, following Palatini \cite{pala}, we see that the Einstein--Hilbert
Lagrangian composed by the Ricci tensor, depending on the
connection\footnote{In this action we concern for the symmetric connection
after the transition to the world indices (see \cite{text}).}, and the
auxiliary field of metric tensor\footnote{We follow the presentation given in
\cite{schroed}.}, which is ordinary defined as a quadratic form of tetrad. Then
the variation of action over the connection leads to the equations of motion
giving the constraints for the covariant derivative of metric tensor, so that
this derivative is equal to zero. This fact implies that the connection is
consistent with the metrics, and it is expressed in terms of Christoffel
symbols. The action is independent of auxiliary field, that leads to the
Einstein--Hilbert equations in the theory of general relativity, so that the
field of metric tensor determining the connection brings the dynamical
characteristics of connection. The procedure of canonical quantization of
Hamiltonian dynamics shows that for this field we have two dynamical massless
modes with the helicity equal to $\pm 2$ \cite{fadd}.

Thus, the vacuum expectation values for the fields $a$ and $b$ defined above
are related with the Planck mass $m_{\rm\scriptscriptstyle Pl}^2 =
v_b^2-v_a^2$,
so that in the case of comparable values\footnote{We argue for this assumption
below.} of $v_a$ and $v_b$ at the energies of dynamical fields less than the
Planck scale we can neglect the contributions quadratic over the strength
tensor
in comparison with the linear Einstein--Hilbert Lagrangian of (\ref{gravity}).
Generally, at $v_b < v_a$ we could get the negative gravitational constant,
i.e. anti-gravity.

Then, we arrive to the Einstein--Hilbert theory of gravity as the low-energy
limit of full Lagrangian with the gauge field condensation, {\it viz.} the
spontaneous breaking of gauge symmtery, so that the vacuum fields lead to the
curved space-time.

\subsection{Cosmological term}
Let us consider the term quadratic over the background vacuum strength in the
action of spin-connection. The corresponding contribution to the Lagrangian
determines the density of vacuum energy due to the background connection. We
can easily show that it is equal to
\begin{equation}
{\cal L}_{[{\cal R}^0]^2} = 3\, g_{\rm\scriptscriptstyle Pl}^2\,
(v_b^2-v_a^2)^2 = 3\, g_{\rm\scriptscriptstyle Pl}^2\,
m_{\rm\scriptscriptstyle Pl}^4,
\label{r2}
\end{equation}
i.e. we get a huge cosmological constant\footnote{It would be quite naively to
think that the expression for the density of vacuum energy in terms of fields
$a$ and $b$ in eq. (\ref{r2}) is the form of potential $V$, if we put
$b^2-a^2=s^2$, then $V\sim - s^4$ and it is not restricted from below. Indeed,
in the quantum chromodynamics, for example, the density of vacuum energy is
negative and proportional to the gluon condensate, so that vacuum energy
infinitely drops with the increase of gluon condensate. The expression for the
density of energy in terms of vacuum fields is generally determined by the
dimensional analysis, and it does not contain a full information on the
nonlinear dynamics, wherein the vacuum expectation values are restricted, and
there is the anomaly in the trace of energy-momentum tensor.}. However, this
contribution could be cancelled by the term determining the interaction of
vacuum field with the spinor current, if there is a corresponding vacuum
condensate of spinor field. Indeed, introduce the vacuum field of spinor
current
\begin{equation}
j_0^{\mu,mn} = \frac{1}{\sqrt{2}}\, \zeta\; \epsilon^{\mu\nu mn} a_{\nu}
\frac{(v_b^2-v_a^2)^2}{v_a^2},
\end{equation}
where $\zeta$ is a dimensionless factor, and calculate the product of current
and the vacuum connection in (\ref{vac}), where the component of connection
with the field $a$ contributes only, so that
\begin{equation}
{\cal L}_{j{\cal R}^0} = - 3\,\zeta\, g_{\rm\scriptscriptstyle Pl}^2\,
(v_b^2-v_a^2)^2.
\label{jr}
\end{equation}
We see that the cosmological constant is cancelled in the sum of quadratic
vacuum curvature (\ref{r2}) and contribution from the interaction of vacuum
spinor and gauge fields (\ref{jr}), if $\zeta=1$. 

Sure, the correlation of vacuum gauge and spinor fields, i.e. ``fine tuning''
necessary for the strict cancellation of cosmological constant, points to some
physical reasons, which nature can be caused by the supersymmetry, for
instance, though that requires an additional study beyond the scope of this
work. We will make the only remark on this problem in section 5. Here we point
out that the strict cancellation of cosmological term can be broken by small
fluctuations of vacuum fields, so that the balance state of system would be
destroyed, and at the appropriate sign of cosmological constant producing 
the repulsion, for example, the inflation expansion of universe would happen
\cite{linde}. In this way, there is a problem of border regions, where one
should match global fluctuations different by their value or, probably, sign,
so that the corresponding domain walls appear not only between two expanding
universes, but also between contracting and expanding ones. 

Thus, we draw the conclusion that the cancellation of cosmological constant can
take place, while the mechanism is not clear enough and should be investigated,
and the fluctuations of vacuum fields result in cosmological implications.

\subsection{Massive modes}
Consider excitations on the background of vacuum field $a$, i.e. in eq.
(\ref{vac}) we substitute for $a^\mu$ by the field $$a^\mu +
\frac{g_{\rm\scriptscriptstyle
Pl}}{\sqrt{2}}\, \omega^\mu(x).$$ 
Then the term of Lagrangian quadratic in $\omega^\mu$ has the form
\begin{equation}
\begin{array}{rcl}
{\cal L}_{\omega} &=&  -\frac{1}{2} \left( (\partial^\mu \omega_\nu)^2
+\frac{1}{2} (\partial^\mu\omega_\mu)^2\right)+ {6}\,{g_{\rm\scriptscriptstyle
Pl}^2}\cdot [3 v_a^2-v_b^2]\, \omega_\mu^2,
\end{array}
\end{equation}
where the term $(\partial^\mu\omega_\mu)^2$ corresponds to the gauge condition,
while the mass of vector field $\omega_\mu$ is determined by the equality
$$
m_\omega^2 = 12\, {g_{\rm\scriptscriptstyle Pl}^2}\, (2v_a^2 - 
m_{\rm\scriptscriptstyle Pl}^2),
$$
so that the necessary constraint for the positive definiteness of both the
square of
$\omega^\mu$ mass and the gravitation constant is the inequality
$$
\frac {1}{3}\, v_b^2 < v_a^2 < v_b^2.
$$
Particularly, the situation of $ v_b^2 - v_a^2 = m_{\rm\scriptscriptstyle Pl}^2
\ll v_a^2 \sim  m_{\rm\scriptscriptstyle Pl}^2 / g_{\rm\scriptscriptstyle
Pl}^2$ is of interest. In this case the vector field has the mass close to the
Planck scale.

Further, let us consider the possibility for a variation of absolute value for
$b$ and introduce 
$$
b^\mu \tilde \phi,
$$
in (\ref{vac}). Then the term of Lagrangian quadratic in $\tilde \phi$ is equal
to
\begin{equation}
{\cal L}_{\phi} = \frac{9}{8} v_b^2 (\partial_\mu \tilde\phi)^2 - 6\,
{g_{\rm\scriptscriptstyle Pl}^2}\,v_a^2 v_b^2\, \tilde\phi^2 
= \frac{1}{2} (\partial_\mu \phi)^2 - \frac{8}{3}\,
{g_{\rm\scriptscriptstyle Pl}^2}\, v_a^2\, \phi^2 ,
\end{equation}
where we have denoted $\phi = {\frac{3}{2} v_b}\, \tilde\phi$, and its mass
equals
$$
m_\phi^2 = \frac{16}{3}\,{g_{\rm\scriptscriptstyle Pl}^2}\, v_a^2.
$$
We see that the mass of scalar field is also at the Planck scale. The above
consideration on the mass of $\phi$ is valid in the vicinity of $\phi\to 0$.
However, we study the situation, when $\langle \phi\rangle = \frac{3}{2} v_b$,
i.e. the field is, probably, displaced from its local minimum, that is an
ordinary scenario with the inflation \cite{linde}.

Summarizing the consideration of massive modes and gravitation Lagrangian, we
describe 3 massive modes of vector field, 1 massive scalar field and 2
polarizations of massless graviton, i.e. 6 degrees of freedom, while in the
standard approach to the gauge theory we should expect that the dynamics
involves 6 kinds of massless particles with two transversal polarizations
because of 6 generators of group transformations, i.e. 12 physical modes. In
addition, we can easily see that the introduction of scalar field for the scale
variation of $a$ analogous to $\phi$ as well as the vector field linear to $b$
similar to $\omega^\mu$ would result in negative kinetic energies for these
additional fields. This fact challenges a more accurate and strict
consideration of spin-connection dynamics, that will be the subject of next
section.

\subsection{Hamiltonian dynamics}
Let us consider the standard approach for the description of gauge field
dynamics. In this way, the field ${\cal A}_{0,mn}$ has no derivative with
respect to time in the Lagrangian, and it is not a dynamical variable. This
field is a Lagrange factor, so that we can put
\begin{equation}
{\cal A}_{0,mn} = 0.
\label{mult}
\end{equation}
Introduce the notations for the components of strength tensor
\begin{equation}
\begin{array}{rcl}
{\cal F}_{0k,0n} = {\cal E}_{k,n}, && {\cal F}_{0k,ij}\cdot\frac{1}{2}\,
\epsilon^{ijn} = \tilde{\cal E}_{k}^{~n},\\[4mm]
{\cal F}_{lp,0n}\cdot \frac{1}{2}\epsilon^{lpk} = {\cal H}^{k}_{~n}, && {\cal
F}_{lp,ij}\cdot\frac{1}{2}\, \epsilon^{ijn} \, \frac{1}{2}\epsilon^{lpk} =
\tilde{\cal H}^{k,n},
\end{array}
\end{equation}
where the indices are running in the limits $(1,2,3)$, and ${\cal E}$ is an
electric field, $\tilde{\cal E}$ is a pseudoelectric field, ${\cal H}$ is a
magnetic field, $\tilde{\cal H}$ is a pseudomagnetic field. Taking into account
(\ref{mult}), we get 
\begin{equation}
{\cal F}_{0i,mn} = \partial_0\,{\cal A}_{i,mn},
\end{equation}
so that the electric fields are given by the velocities of gauge fields.

Calculating the trace for the square of strength tensor and the product of
strength tensor with the tensor dual over the space-time world indices,
\begin{equation}
{\cal F}^{D}_{\mu\nu,mn} = \frac{1}{2}\, \epsilon_{\mu\nu}^{~~\alpha\beta} \,
{\cal F}_{\alpha\beta,mn},\;\;\;\;\;
{\cal F}_{\mu\nu,mn} = -\frac{1}{2}\, \epsilon_{\mu\nu}^{~~\alpha\beta} \,
{\cal F}^{D}_{\alpha\beta,mn},
\label{du}
\end{equation}
we find the following expressions for the field invariants, which are scalar
and do not depend on the gauge:
\begin{equation}
\begin{array}{rcl}
I_1 &=& {\cal E}\cdot {\cal H} - \tilde{\cal E}\cdot \tilde{\cal H}, \\[3mm]
I_2 &=& {\cal E}\cdot \tilde{\cal E} - {\cal H}\cdot \tilde{\cal H}, \\[3mm]
I_3 &=& {\cal E}\cdot \tilde{\cal H} + \tilde{\cal E}\cdot {\cal H}, 
\end{array}
\end{equation}
along with the Lagrangian of gauge field
\begin{equation}
{\cal L} = \frac{1}{2 g_{\rm\scriptscriptstyle Pl}^2}\, \left[{\cal E}^2 -
\tilde{\cal E}^2 - {\cal H}^2 + \tilde{\cal H}^2\right],
\label{l}
\end{equation}
where the scalar products are contracted over the three-dimensional Euclidean
indices in the group and space.

In the formalism under consideration the momenta of fields coincide with the
electric fields. The Hamiltonian of system can be written down in the form
\begin{equation}
{\sf H} = \frac{1}{2 g_{\rm\scriptscriptstyle Pl}^2}\, \left[ {\cal E}^2 -
\tilde{\cal E}^2 + {\cal H}^2 - \tilde{\cal H}^2\right],
\label{pt}
\end{equation}
so that introducing the gauge
\begin{equation}
\partial_i\, {\cal A}_{i,mn} = 0,
\end{equation}
we can calculate its Poisson bracket with the Hamiltonian of (\ref{pt}), i.e.
the time derivative of constraint, and get new constraint (the Gauss law)
\begin{equation}
\partial_i\, \partial_0\, {\cal A}_{i,mn} = 0,\;\; \Leftrightarrow\;\;
\partial_i\, {\cal E}_{i,n} = 0,\;\;\; \partial_i\, \tilde{\cal E}_{i,n} =0.
\end{equation}
Thus, Hamiltonian (\ref{pt}) depends on the transverse components of electric
fields as well as the magnetic fields expressed in terms of transverse gauge
fields, so that
\begin{equation}
{\sf H} = \frac{1}{2 g_{\rm\scriptscriptstyle Pl}^2}\, \left[ {\cal
E}_{\perp}^2({\cal A}_\perp) - \tilde{\cal E}_{\perp}^2({\cal A}_\perp) + {\cal
H}^2({\cal A}_\perp) - \tilde{\cal H}^2({\cal A}_\perp)\right].
\label{ptperp}
\end{equation}

We know one exact solution of field equations following from the Hamiltonian of
(\ref{ptperp}): all of the fields and momenta are equal to zero.
However, we see in (\ref{ptperp}) that such the Hamiltonian results in the
instability of zero solution, and the perturbation theory built in the vicinity
of zero includes the modes with the negative kinetic energy (pseudoelectric
field). Nevertheless, this fact does not imply that there is no stable state in
the theory since the nonlinear character of field equations can generally lead
to a restriction of total energy from below. It is evident that in the
case of existing the stable solution it is determined by dynamical fields,
which kinetic energy could differ from zero.

A possibility of such situation can be shown in the following way. For the
fields determined by the dual strength tensor we have the relations
\begin{equation}
\tilde{\cal E}_k^{~n} = - [\tilde{\cal H}^D]_{k}^{~n},\;\;\;\;
\tilde{\cal H}^{kn} = - [\tilde{\cal E}^D]^{kn}.
\end{equation}
Therefore the Lagrangian of (\ref{l}) can be rewritten in the form 
\begin{equation}
{\cal L} = \frac{1}{2 g_{\rm\scriptscriptstyle Pl}^2}\, \left[{\cal E}^2 -
[\tilde{\cal H}^D]^2 - {\cal H}^2 + [\tilde{\cal E}^D]^2\right],
\label{ld}
\end{equation}
so that eq. (\ref{ld}) leads to the  Hamiltonian
\begin{equation}
{\sf H} = \frac{1}{2 g_{\rm\scriptscriptstyle Pl}^2}\, \left[ {\cal
E}_{\perp}^2({\cal A}_\perp) + [\tilde{\cal E}^D_{\perp}]^2(\tilde{\cal
A}^D_\perp) + {\cal H}^2({\cal A}_\perp,\tilde{\cal A}^D_\perp) + [\tilde{\cal
H}^D]^2({\cal A}_\perp,\tilde{\cal A}^D_\perp)\right],
\label{ptperpd}
\end{equation}
where we have introduced the fields\footnote{Under the conditions of
transversity for the fields of (\ref{dfield}) we can show that the Poisson
bracket $\{\tilde{\cal H},{\cal E}\} = -\{\tilde{\cal
E}^D,{\cal E}\}\approx 0$ on the surface of constraints, i.e. the momenta of
fields ${\cal A}$ and $\tilde{\cal A}^D$ commute. The fact that the fields
$\tilde{\cal H}^D$ and $\cal H$ do not depend on the time derivatives, follows
from the equations determining the exclusion of fields $\tilde{\cal A}$ and
${\cal A}^D$: $$\left\{\begin{array}{lcl} \partial_0\,
\tilde{\cal A} &=& - \tilde{\cal H}^D({\cal A}^D,\tilde{\cal A}^D),\\
\partial_0\, {\cal A}^D &=& - {\cal H}({\cal A},\tilde{\cal
A}).\end{array}\right. $$ Thus, it is possible that the dependence of magnetic
fields on ${\cal A}$ and $\tilde{\cal A}^D$ becomes nonlocal in time (this
should not break the causality, since we transform the theory, wherein the
causality principle is valid). However, we see that the expression of
Lagrangian in terms of non-homogeneous fields ${\cal A}$ and $\tilde{\cal A}^D$
generally leads to the problem on the separation of variables.}
\begin{equation}
\begin{array}{lcl}
{\cal A}_{\perp i,n} = {\cal A}_{i,0n}, && \partial_i\, {\cal
A}_{i,0n}=0,\\[3mm]
\tilde{\cal A}^D_{\perp i,n} = \frac{1}{2}\,\epsilon_{kln} {\cal
A}^D_{i,kl}, && \partial_i\, {\cal A}^D_{i,kl}=0,\\[3mm]
\partial_i {\cal E}_{\perp i,n} = 0, && \partial_i \tilde{\cal E}^D_{\perp i,n}
= 0,\\[3mm]
{\cal E}_{\perp i, n} = \partial_0\,{\cal A}_{\perp i,0n}, && 
\tilde{\cal E}^D_{\perp i, n} = \frac{1}{2}\,\epsilon_{kln}\,
\partial_0\,{\cal A}^D_{\perp i,kl},
\end{array}
\label{dfield}
\end{equation}
under the condition that the gauge field allows such the separation of
variables, of course, and we can believe that putting the Lagrange factors
equal to zero is consistent
$$
{\cal A}_{0,mn} = 0,\;\; 
\;\; \tilde{\cal A}^D_{0,mn} = 0.
$$
The expression for the Hamiltonian in (\ref{ptperpd}) shows its positive
definiteness. However, the construction of perturbation theory under this
Hamiltonian is problematic since we assume the implicit substitution of
particular gauge fields in terms of dual field, while we must make explicit
substitutions in the nonabelian vertices of interactions. Nevertheless, we
suggest that the operation with the dual fields allows us to present some
characteristic features of spin-connection gauge field\footnote{Sure, in this
construction we suppose the assumption on a possibility of correct separating
the variables.}.

\subsubsection{Dual ansatz}
Following the notations of (\ref{vac}) in section 3.2, introduce the dual
strength tensor as a sum over the tensors for the components of connection
${\cal A^D}$, so that
\begin{equation}
{\cal F}^D_{\mu\nu, mn} = {\cal R}^{0[a^D]}_{\mu\nu, mn} + {\cal
R}^{0[b^D]}_{\mu\nu, mn} + \frac{1}{2}{\cal R}_{\mu\nu, mn}(\Gamma^D),
\end{equation}
and 
\begin{eqnarray}
{\cal A}^{[b^D]}_{\mu,mn} &=& \frac{1}{\sqrt{2}}\, \epsilon_{\mu\nu mn}
[b^D]^\nu ,\nonumber\\
{\cal A}^{[a^D]}_{\mu,mn} &=& \frac{1}{\sqrt{2}}\, (\eta_{\mu m} a^D_n -
\eta_{\mu n} a^D_m), \label{vacd}\\ 
{\cal A}^{[\Gamma^D]}_{\mu,mn} &=& \frac{1}{2}\, \Gamma^D_{\mu,mn},
\end{eqnarray}
where the vacuum fields $a^D$ and $b^D$ have the expectation values
\begin{equation}
\begin{array}{rcl}
\langle a^D_\mu\rangle = 0,\;\;\; \langle b^D_\mu\rangle = 0,\;\;\; 
&&\langle a^D_\mu a^D_\nu \rangle = \frac{1}{4}\, \eta_{\mu\nu}
g^2_{\rm\scriptscriptstyle Pl} [v^D_a]^2,\;\;\; \\[4mm] 
\langle a^D_\mu b^D_\nu \rangle = \frac{1}{4}\, \eta_{\mu\nu}
g^2_{\rm\scriptscriptstyle Pl} v^D_a \, v^D_b,\;\;\; &&
\langle b^D_\mu b^D_\nu \rangle = \frac{1}{4}\, \eta_{\mu\nu}
g^2_{\rm\scriptscriptstyle Pl} [v^D_b]^2 .
\end{array}
\end{equation}
Then, we have the following contribution to the vacuum strength tensor:
\begin{equation}
\begin{array}{rcl}
-\frac{1}{2}\, \epsilon_{\mu\nu}^{~~\alpha\beta}\,{\cal
R}^{0[b^D]}_{\alpha\beta, mn}(b^D) &=&
- (b_D^2\, \epsilon_{\mu\nu mn} - \epsilon_{\mu\nu m\alpha}\, b_D^{\alpha}
b^D_n + \epsilon_{\mu\nu n\alpha}\, b_D^{\alpha} b^D_m),\\[3mm]
{\cal R}^{0[a^D]}_{\mu\nu, mn}(a^D) &=& - {\cal R}^{0[b^D]}_{\mu\nu, mn}(a^D) .
\end{array}
\end{equation}
Noting
$$
{\cal F}^2 = -{\cal F}_D^2,
$$
we see that the term of Lagrangian linear over the background strength tensor
and the dynamical tensor of curvature is equal to\footnote{We do not consider
terms in the form of $(v_a^2-v_b^2) \epsilon^{\mu\nu mn}{\cal R}_{\mu\nu
mn}(\Gamma^D)$ as well as that of dual to it, since such contributions are
equal to zero for the symmetric connection with the indices on the world
coordinates.}
\begin{equation}
\begin{array}{rcl}
{\cal L}_G &=&  - \frac{1}{4}\, (v_b^2-v_a^2)\; {\cal R}(\Gamma)-\frac{1}{4}\,
([v^D_b]^2-[v^D_a]^2)\; {\cal R}(\Gamma^D).
\end{array}
\label{gravityd}
\end{equation}
It is evident that the motion equations for the connections $\Gamma$ and
$\Gamma^D$ lead to zero covariant derivatives of metric tensor over both
connections, so that in the linear limit over the strength of dynamical
connection-fields the massless modes are coherent, and the connections are
expressed by the symbols of Christoffel, $\Gamma^D=\Gamma$ and $R(\Gamma^D) =
R(\Gamma)$. Therefore
\begin{equation}
{\cal L}_G =  -\frac{1}{4}\, \left(v_b^2-v_a^2+[v^D_b]^2-[v^D_a]^2\right)\;
{\cal R}(\Gamma).
\label{dualgravity}
\end{equation}
It is interesting to note that the cosmological constant determined by the
vacuum fields in the strength tensor and dual one takes the form
\begin{equation}
{\cal L}_{[{\cal R}^0]^2} = 3\, g_{\rm\scriptscriptstyle Pl}^2\,
[(v_b^2-v_a^2)^2 - ([v^D_b]^2-[v^D_a]^2)^2] ,
\label{r2du}
\end{equation}
and it is cancelled, if\footnote{In comparison with section 3.2, the relation
between the difference of vacuum expectation values and the Planck mass
changes.}
\begin{equation}
|[v^D_b]^2-[v^D_a]^2| = |v_b^2-v_a^2| = \frac{1}{2}\, m_{\rm\scriptscriptstyle
Pl}^2.
\label{cosmo}
\end{equation}
Thus, the cancellation of cosmological term in this mechanism does not require
an introduction of spinor condensate, but we can again expect that fluctuation
of vacuum fields probably could lead to the expanding or contracting universe.

The expectation value of vacuum strength is given by the expression
\begin{equation}
\begin{array}{lcl}
\langle{\cal F}_{\mu\nu,mn}\rangle = \frac{1}{2}\, (\eta_{\mu m}\eta_{\nu n} -
\eta_{\mu n}\eta_{\nu m})\, (v_a^2-v_b^2) - \frac{1}{2}\,\epsilon_{\mu\nu mn}
([v^D_b]^2-[v^D_a]^2),
\label{vevF}
\end{array}
\end{equation}
so that the expectation values of electric and magnetic
fields\footnote{Calculating the squares of fields entering the Lagrangian, we
take into account their nonzero dispersion.} are equal to
\begin{equation}
\begin{array}{lcl}
\langle{\cal E}_{m,n}\rangle = \frac{1}{2}\, (v_a^2-v_b^2)\delta_{mn}, &&
\langle{\cal H}_{m,n}\rangle = \frac{1}{2}\, ([v^D_a]^2-[v^D_b]^2)\delta_{mn},
\\[4mm] \langle\tilde{\cal H}_{m,n}\rangle = \frac{1}{2}\,
(v_a^2-v_b^2)\delta_{mn}, && \langle\tilde{\cal E}_{m,n}\rangle = \frac{1}{2}\,
([v^D_a]^2-[v^D_b]^2)\delta_{mn}.
\end{array}
\end{equation}

The consideration of massive modes is analogous to section 3.4, and it leads to
massive vector and scalar fields\footnote{The cross-terms linear in the massive
fields and dual strength tensor of vacuum fields do not give new effects: the
gauge of vector fields takes a general form of $\partial^\mu\, \omega_\mu -
\Phi(a^d,b^d) = 0$, and for the scalar fields the terms three-linear in the
vacuum field are equal to zero.}, so that
\begin{equation}
\begin{array}{lcl}
~\,m_\omega^2 = 12\, {g_{\rm\scriptscriptstyle Pl}^2}\, \left( 2 v_a^2 -
\frac{1}{2}\,m_{\rm\scriptscriptstyle Pl}^2\right), &&
~\,m_\phi^2 = \frac{16}{3}\,{g_{\rm\scriptscriptstyle Pl}^2}\, v_a^2, \\[5mm]
m_{\omega_D}^2 = 12\, {g_{\rm\scriptscriptstyle Pl}^2}\, \left( 2 [v^D_a]^2 -
\frac{1}{2}\,m_{\rm\scriptscriptstyle Pl}^2\right), &&
m_{\phi_D}^2 = \frac{16}{3}\,{g_{\rm\scriptscriptstyle Pl}^2}\, [v^D_a]^2.
\end{array}
\end{equation}
We see that the number of gauge constraints for the fields is equal to 6
because\footnote{We put $\tilde\omega_\mu=b_\mu\tilde\phi$.}
$$
\begin{array}{lcl}
a^2- v_a^2 = 0, && b^2- v_b^2 = 0, \\[4mm]
\partial^\mu\, \omega_\mu = 0 && \tilde\omega_\mu - b_\mu\, (b\cdot
\tilde\omega) = 0,
\end{array}
$$
while the analogous constraints for the fields determining the dual strength
tensor probably represent the additional conditions for the momenta of gauge
fields, since in the dual notations they are transformed in some ordinary
constraints for the fields.

Thus, we have got 12 dynamical fields: 3 polarizations of massive vector field
and 3 ones dual to them, the scalar massive field and that of dual to it, as
well as 2 modes of massless field with the spin 2 and 2 dual modes, which are
coherent in the limit of Einstein gravity, since the massless modes differ at
the level of corrections suppressed by the square of ratio given by the energy
of field quanta over the Planck mass.

\subsubsection{Vacuum fields and equations of motion}

In the previous sections we do not address the problem on how one could
actually show an effective action providing the above choice of background
vacuum fields, otherwise the prescriptions (\ref{vacF}), (\ref{vac}) look
rather random. Of course, such the presentation of effective action would imply
an exact or, at the very least, approximate solution of field equations in a
manifest form, that seems to be quite a difficult task as concerns for
nonabelian theories. The similar situation is observed in QCD, where we know
that the gauge field has a vacuum condensate involving an infrared
nonperturbative dynamics, however, we cannot prove or derive a form of
effective action except its modelling. In the theory under study we assume the
appearance of gauge field condensation, in general, and provide the expressions
modelling this phenomenon. The form of vacuum field ansatz is quite transparent
from the Lorentz-structure point of view. However, the formulae cannot be
proved to the moment. 

In this section we investigate, how the vacuum fields enter the field
equations. First, we suggest that the vacuum expectation of strength tensor
presented above is valid in the Fock-Schwinger gauge of fixed point
$$
x^\mu\cdot {\cal A}_{\mu, mn}(x) = 0,
$$
that implies the introduction of special marked point in the configuration
space. In this gauge the field can be expressed in terms of the strength tensor
\begin{equation}
{\cal A}_{\mu, mn}(x) = \int\limits_0^1 dz\, z\;x^\nu\cdot {\cal F}_{\nu\mu,
mn}(z\cdot x),
\label{FS}
\end{equation}
and we put the vacuum expectation $\langle {\cal A}_{\mu, mn}(x) \rangle$ by
making use of (\ref{vevF}) in the right-hand side of (\ref{FS}) and integrating
out since the vacuum strength tensor is independent of $x$. However, the form
(\ref{FS}) of Fock-Schwinger gauge in the nonabelian theory is not
consistent with the global vacuum expectation of strength tensor. Formally, we
could joint (\ref{vevF}) and (\ref{FS}) with no contradiction, i.e. we could
reproduce the strength tensor by its definition under the substitution of
(\ref{FS}) if we ignore the totally antisymmetric tensor
$\epsilon_{\mu\nu\alpha\beta}$ multiplied by the mixed product of vacuum
expectations for the gauge field and dual to it and suppose (\ref{cosmo}), that
implies the cancellation of cosmological term. Moreover, the vacuum strength
tensor under the condition of (\ref{cosmo}) is anti-selfdual, and hence, it
satisfies the field equations with no sources. This selfduality, as well known,
results in the local minimization of gauge field action.

In detail, the field equations are given by
$$
[\nabla^\mu({\cal A}) {\cal F}_{\mu\nu}(x)]_{mn} = j_{\nu, mn}(x),
$$
where the covariant derivative is defined in section \ref{3.1}. For the vacuum
field without a dispersion we put
$$
[\nabla^\mu(\langle{\cal A}\rangle)\; \langle{\cal F}_{\mu\nu}(x)\rangle]_{mn}
=
\langle j_{\nu, mn}(x)\rangle,
$$
which under (\ref{FS}) straightforwardly results in
\begin{eqnarray}
\langle j^{\alpha,mn} \rangle&=& \frac{1}{2}\left((v_a^2-v_b^2)^2 -
        \left([v_a^D]^2-[v_b^D]^2\right)^2\right)
      ({g^{m \beta }}{g^{n \alpha }}-
        {g^{m \alpha }}{g^{n \beta }})x_\beta \nonumber\\ &-&
     (v_a^2-v_b^2) ([v_a^D]^2-[v_b^D]^2)
      \epsilon^{m n \alpha  \beta }x_\beta.
\end{eqnarray}
As we have already mentioned, the first term in the above equation disappears
under the anti-selfduality and the cancellation of cosmological constant, while
the second is the artefact of (\ref{FS}), and it should be ignored, which is
evident since the selfduality of strength tensor guarantees that the source
current is equal to zero.

Therefore, the field equations are satisfied until
\begin{enumerate}
\item
the cosmological constant is put to zero;
\item
the spinor current is equal to zero.
\end{enumerate}

Thus, we have shown how the offered vacuum condensates can enter the field
equations and be related with the physical contents of the model under
consideration.

\section{Dirac spinors}
The Dirac bispinors
$$\psi = \left( {\theta_\alpha}\atop {\bar\chi^{\dot\alpha}}\right), $$
are operated by the algebra of $\gamma$-matrices, which are defined by the
following relations:
\begin{equation}
\gamma^m = \rho_+ \raisebox{1.7pt}{$\scriptscriptstyle \boldsymbol \otimes$}
\sigma^m + \rho_- \raisebox{1.7pt}{$\scriptscriptstyle \boldsymbol
\otimes$}\bar\sigma^m, \end{equation}
so that
\begin{equation}
\{\gamma^m,\gamma^n\} = 2\,\eta^{mn},
\end{equation}
because $\rho_{+}=\frac{1}{2}(\sigma_1+i\,\sigma_2)$ is a raising operator,
$\rho_- =\rho_+^\dagger$ and
\begin{equation}
\rho_+\rho_- + \rho_-\rho_+ = 1,\;\;\;\; \rho_+^2 =0. 
\label{proj}
\end{equation}
Eq. (\ref{proj}) points to that the quantities $\rho_L = \rho_+\rho_-$,
$\rho_R=\rho_-\rho_+$ are projectors. Then we find
\begin{equation}
\{\gamma^n,\gamma^m\} = \rho_+\rho_-\raisebox{1.7pt}{$\scriptscriptstyle
\boldsymbol \otimes$}\{\sigma^n\bar\sigma^m
+\sigma^m\bar\sigma^n\} + \rho_-\rho_+\raisebox{1.7pt}{$\scriptscriptstyle
\boldsymbol \otimes$}\{\bar\sigma^n\sigma^m +\bar\sigma^m\sigma^n\}
=2\,\eta^{nm}.
\end{equation}

\subsection{Internal symmetry}
Since the Weyl factors enter the definitions of Dirac $\gamma$-matrices and
bispinors, the above study of invariant representation for the
$\sigma$-matrices remains valid. The additional subject is the invariant
changes of representations for the matrices $\rho$:
\begin{equation}
\rho_+^\prime = e^{i\lambda}\, f\cdot \rho_+\cdot f^{-1},
\end{equation}
where $f\in SU(2)$, $\lambda\in {\mathfrak R}$ can be local functions. As the
left-handed and right-handed spinors in the massless Dirac field, which get the
mass due to the interaction with scalar particles, are independent and they can
be separately put equal to zero, the additional group of symmetry for the Dirac
spinors is the product of groups on the chiral fields, so that
\begin{equation}
U(1)_L\raisebox{1.7pt}{$\scriptscriptstyle \boldsymbol
\otimes$}SU(2)_L\raisebox{1.7pt}{$\scriptscriptstyle \boldsymbol
\otimes$}U(1)_R\raisebox{1.7pt}{$\scriptscriptstyle \boldsymbol
\otimes$}SU(2)_R.
\label{gD}
\end{equation}
This fact can be verified and confirmed by a straightforward calculation, if we
write down the generators of unitary transformations for representations of
Dirac spinors in the form of sum over 16 independent basis matrices and find
the constraints for the coefficients of linear span from the equation for the
conservation of Hamiltonian form for the Dirac field in the massless case.

The group of internal symmetry (\ref{gD}) includes the group of electroweak
symmetry, which is $U(1)\raisebox{1.7pt}{$\scriptscriptstyle \boldsymbol
\otimes$}SU(2)_L$. We see that the combining of two Weyl spinors in the
multiplet of bispinor involves the gauge symmetry for the Dirac spinors, so
that, if we identify the group of electroweak symmetry with the subgroup of
(\ref{gD}), then we expect that the partner of electron in the weak
iso-doublet, i.e. neutrino, should be also the Dirac particle. A trivial
speculation concerns for that the group of internal symmetry could be extended,
if the multiplet is built by several Weyl spinors, and the group of quantum
chromodynamics, for instance, does not contradict with the above investigation
for the Dirac spinors. More interesting challenge is an observability of
extension for the electroweak group, that follows from (\ref{gD}). So, the
multiplet can be really more extended than the Dirac spinor, and hence, the
additional group includes the subgroup $U(1)\raisebox{1.7pt}
{$\scriptscriptstyle \boldsymbol \otimes$} SU(2)_R$.

Thus, in contrast to the Weyl spinors, for the invariant representations of
Dirac algebra on the bispinors there is the gauge group of internal symmetry
(\ref{gD}) including the electroweak group.

\section{Supersymmetry}
In the study of possible mechanism for the cancellation of cosmological
constant in section 3.3, the problem on the necessity of supersymmetry between
the bosonic and fermionic fields has been mentioned.

In this section we discuss the problem on supersymmetry from a general point of
view and offer a BRST-generalization of operator for the evolution of Weyl
spinor, that leads to the supersymmetric transformations in the space of world
coordinates and global spinors. Thus, the introduction of parametric
transformations determined by global Grassmann variables for the Lagrangian of
Weyl spinors results in the necessity of supersymmetry in the theory with the
Weyl spinors.

For the Lagrangian of Weyl spinor
$$
{\cal L} = \theta \sigma^\mu p_\mu \bar \theta,
$$
the partition function\footnote{The partition functional of Green functions can
be represented as the limit of partition function product: $\lim
\limits_{M\to\infty}\prod \limits_{m=1}^{M} \int d\theta_m d\bar \theta_m
\; e^{i \theta_m \sigma^\mu p_\mu \bar \theta_m d x_m +i\eta_m \theta_m d x_m 
+i\bar\theta_m \bar \eta_m d x_m}$, where we can redefine the fields $\theta_m
=\theta(x_m)$ and $\eta_m=\eta(x_m)$, so that the element of space-time $d x_m$
would enter in a implicit way, that allows us to make a transition to global
spinors in a simple manner. Therefore, we can consider the partition function
for our purpose.}
$W$ depending on sources $\eta,\; \bar \eta$, has the form
$$
W(\eta,\bar \eta)=\int d\theta d\bar \theta \; e^{i{\cal L}+i\eta \theta
+i\bar\theta \bar \eta}= N_W\cdot e^{iG}=N_W\cdot \exp\{-i \bar \eta \bar K
\eta\},
$$
where
$$
\bar K \cdot \sigma^\mu p_\mu = 1,
$$
so that the fields are equal to
$$
\theta= -i\frac{1}{W}\frac{\partial W}{\partial \eta}= \frac{\partial G}
{\partial \eta},\;\;\;\;
\bar \theta= -i\frac{1}{W} \frac{\partial W}{\partial \bar \eta}=
\frac{\partial
G} {\partial \bar \eta}.
$$

It is convenient to fix the normalization of fields so that ${\cal L}$ is the
dimensionless quantity, and the Weyl spinor has the dimension of inverse square
root of energy (we can easily reconstruct the usual normalization of fields and
Lagrangian with no problem). In this way, the element of space-time measure is
included into the normalization.

The operator of evolution has the form
$$
U(d\tau)= e^{i{\cal L}d\tau},
$$
where the field is given by the derivatives over the sources, and $W(U)=U\cdot
W \cdot U^{\dagger}$. In the ordinary normalization we have $d\tau= E dt$, $E$
is an energy of state, $dt$ is a shift of time.

Under the action of operator $U$, the fields on the partition function $G$
get the phase shift $d\tau$, reflecting the arbitrary choice of time
reference-point on the trajectory of free Weyl spinor. Indeed, in accordance
with the Hausdorff formula
$$
e^A\cdot e^B =e^{A+B+[A,B]/2+\ldots}
$$
we find
$$
\delta \theta \cdot G = i[{\cal L}, -\bar \eta \bar K] d\tau \cdot G = -i
d\tau \; \theta \cdot G
$$
so that $\theta(d\tau)=e^{-id\tau} \theta$.

The BRST-generalization of evolution operator can be obtained by fixing the
shift in the form 
$$
d\hat \tau = 2i\left[\frac{\xi \theta}{\theta\theta}-\frac{\bar \xi \bar
\theta}{\bar \theta \bar \theta}\right],
$$
where $\xi$ is an infinitesimal Weyl spinor, the parameter of transformation,
so that $\{\xi,\theta\}=0$ and
$$
\delta_\xi =i{\cal L}d\hat \tau = -\xi \sigma^\mu p_\mu \bar \theta +
\theta \sigma^\mu p_\mu \bar \xi.
$$
Then we find that the action of operator on the fields gives 
\begin{eqnarray}
\delta_\xi \theta\cdot G &=& [\delta_\xi, \frac{\partial G}{\partial
\eta}]\cdot G = [\delta_\xi, - \bar \eta \bar K]\cdot G= \xi \cdot G,\\
\delta_\xi \bar \theta\cdot G &=& [\delta_\xi, \frac{\partial G}{\partial \bar
\eta}]\cdot G = [\delta_\xi, - \bar K \eta]\cdot G= \bar\xi \cdot G.
\end{eqnarray}
The local operator of coordinate gets the shift
$$
\delta_\xi x_\mu  \cdot G= (-i\xi \sigma_\mu \bar \theta +
i\theta \sigma_\mu \bar \xi) \cdot G,
$$
and we see that the considered BRST-transformation makes the group of
supersymmetry in the superspace ${\mathfrak R}^{4|4}=\{x_\mu,\theta,\bar
\theta\}$.

Indeed, the action of commutator on the partition function has the form
\[
[\xi Q,\bar Q \bar \xi]\cdot G = 2\xi \sigma^\mu p_\mu \bar \xi,
\]
where we have introduced the notation
\[
\delta_\xi = \xi Q +\bar Q \bar \xi.
\]
Taking into account the anti-commutative properties of transformation
parameters $\xi,\; \bar \xi$, we get
\[
\{Q_\alpha,\bar Q_{\dot \beta}\} = 2 \sigma^\mu_{\alpha\dot \beta} p_\mu .
\]
A construction of Lagrangians invariant under the supertransformations in the
theory of interacting fields can be performed in the ordinary technique of
superfields, wherein the $\theta\theta\cdot \bar \theta \bar \theta$-components
are transformed as full divergences and, hence, produce the invariant
Lagrangians.

\begin{figure}[th]
\centerline{\epsfxsize=7cm \epsfbox{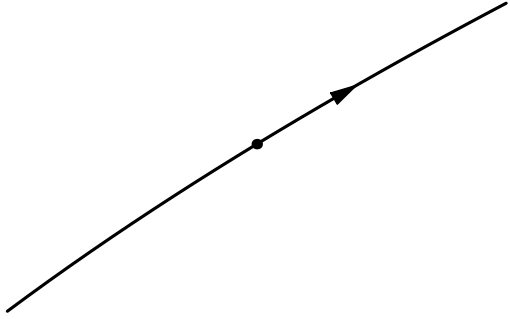}}
\caption{The displacement of zero-point at the trajectory of free spinor.}
\label{traject}
\end{figure}

As for the physical meaning of such the derivation of supertransformations, we
deal with the fact that the choice of reference zero-point at the trajectory of
free spinor parametrized by $\tau$ is conventional, and it does not carry any
measurable information (see Fig.\ref{traject}).

It is rather evident that under the displacement of reference point along the
trajectory the physical observables do not change, i.e. there is an invariance,
while the spinor coordinates and its wave function $\theta$ acquire the
variations
\begin{eqnarray}
\delta_\tau x_\mu &=& v_\mu\, d\tau,\nonumber\\
\delta_\tau \theta &=& -{\rm i}\, d\tau\; \theta .
\label{direct}
\end{eqnarray}
The transformations of supersymmetry solve the inverse problem: at the fixed
variation of spinor wave-function
$$
\delta_\xi \theta = \xi = \delta_\tau \theta 
$$
in accordance with (\ref{direct}) one can find the change of spinor coordinates
depending on the infinitesimal value $\xi$, which implicitly gives the shift of
trajectory parameter $\tau$. We have shown that such the solution can be
constructed in the explicit form implying the introduction of supersymmetric
transformations in the superspace if the spinor is global.

Thus, the supersymmetry has to be considered as necessary ingredient in the
theory of interactions. Items on the structure of supersymmetric multiplets and
spontaneous breaking of supersymmetry need a model study.

\section{Discussion}
The standard model of particle interactions, which finds an increasing number
of its precise experimental confirmations, is built on the principle
of local gauge group acting on the spinors, so that the Lagrangian remains
invariant under the introduction of gauge vector fields. This theory is stable
under the quantum loop corrections, since they do not lead to a necessary
introduction of infinite number of new physical quantities, and the
renormalizability provides the description of quantum fields in terms of their
masses and charges, originally involved in the theory in the formulation at the
classical level. The Einstein--Hilbert theory of gravitation\footnote{See a
modern textbook by W.Siegel \cite{text}.} stands in a separate point, since it
is based on no gauge principle with the vector fields mediating the
interaction, but it operates with the metric tensor fields of spin 2, and in
the calculations of quantum loops the theory of gravity generally needs to
introduce an infinite number of new physical quantities, since it is not
renormalizable. Moreover, the intermediate vector fields lead to a repulsive
force for  charges with the same sign and to a attraction for charges of
different sign, so that trivial attempts to derive the Lagrangian of gravity on
the basis of gauge principle with a group not connected with the space-time
characteristics of particles, are certainly disfavored because the gravity does
not distinguish the charges of sources by their signs. The nonrenormalizability
of Einstein gravity is, first, a consequence of form for the interaction of
space-time metrics with the energy-momentum tensor. Since the dimension of
energy-momentum tensor is equal to four in the energy units, the rescaling of
metrics to the canonical unit dimension of bosonic field\footnote{In the
momentum space of $k$, the propagator of such field behaves as $1/k^2$ with no
other dimensional parameters.} results in that the coupling constant should be
equal to an inverse mass, so that loop corrections get increasing powers of
divergency. Second, after the rescaling to the dimensional bosonic field of
metrics, the factor providing the invariance of space-time volume measure,
$\sqrt{-g}$, has a dimension equal to 2 and involves an appropriate multiplying
factor as a coupling constant of dimension minus 2, that leads to the increase
of divergency powers in the loop calculations with the matter, too. So, the one
loop corrections require the introduction of curvature tensor squared, that
implies the presence of higher space-time derivatives for the metrics as well
as new coupling constants and cosmological term \cite{birreldavis}.
Moreover, the Einstein theory of gravity has intrinsically got an original
scale of energy, that is determined by the dimensional gravitational
constant, while in the gauge theories the presence of intrinsic scale takes
place only in the spontaneous breaking of vacuum symmetry, the condensation of
gauge fields and the renormalization group breaking of conformal invariance. On
the other hand, the fundamental scale as in the gravitation can point to that
there is a dimensional quantity in the primary theory that could be naturally
involved in theories of nonlocal extended objects, for example, in the string
theory. In this respect, the quantum theory of gravity can be constructed with
no connection with the gauge principle and renormalizability. Some implications
of such programme are well represented in modern studies of M-theory \cite{M}.
We emphasize also that the gauge interaction in extended dimensions leads to a
necessary introduction of fundamental scale of energy, since the coupling
constant becomes dimensional, that seems to be not attractive from the logical
point of view because, for example, the consideration of scalar complex field,
i.e. simple operation with the complex numbers, involves a necessary
fundamental length. So, a parting with the powerful features of gauge theories
in four dimensions toward the theories in extended dimensions does not seem to
be crucial. In another respect, a conversion to the theories of extended
objects makes not only a logical step forward a denying the completeness of
local quantum field theory, but also a more wide usage of mathematical language
with all of its beauty and complication of analysis. In the present paper we
have attempted to keep the quantum gravity in the framework of local quantum
field theory by exploring a gauge symmetry possessing some nontrivial
space-time properties and the gauge field condensation implying the spontaneous
breaking of gauge symmtery.

We have studied the equivalence of spinor algebra representations as the gauge
symmetry and found that the corresponding current by Noether for the Weyl
spinors has occurred the current of interaction with the spin-connection. The
global invariance of Lagrangian has been provided by the introduction of
compensating gauge transformation for the auxiliary field of tetrad. The local
gauge group has led to the introduction of nonabelian gauge fields, for which
the invariant Lagrangian has been defined. Then we have considered the example
of nonzero vacuum fields, which have caused the spontaneous breaking of gauge
symmtery due to the condensation and given the effective low-energy
Einstein--Hilbert action of gravity with the massless modes of spin 2, while
the other degrees of freedom for the gauge field have got the masses. We have
shown that the most aesthetic variant of modelling the vacuum structure has
involved the fields determining the dual strength tensor of gauge field, so
that in this way, the mechanism for the cancellation of cosmological constant
has had the most natural form due to the symmetry between the gauge vacuum
fields and those of dual strength tensor. In the other mechanism, the
cancellation of cosmological term has taken place due to the condensate of
spinor field, that has to be tuned with the condensates of gauge fields. This
fact could suggest the underlying supersymmetry between the fermionic and
bosonic fields. We have shown that the necessary introduction of supersymmetry
has followed from the BRST-generalization of evolution operator for the Weyl
spinor, since the physical observables of free Weyl particle have not to depend
on the choice of time reference-point on the trajectory. Further we have shown
which changes have been brought into the theory by investigating the multiplet
of Weyl spinors composing the Dirac bispinor. In this case, the additional
internal gauge symmetry including the standard model electroweak symmetry has
occurred.

The most challenging problems of offered approach are a consistent construction
of perturbation theory, a canonical quantization, a prove of renormalizability,
a derivation of asymptotic freedom\footnote{We can easily find the expression
for the structure constants of group under study from the definition of
covariant derivative in the adjoint representation (\ref{nabla}). So, for
$f^K_{PM}$, where the anti-symmetric group indices are equal to $K=kl$, $P=pq$,
$M=mn$, we get $$f^{kl}_{pq,mn} = \frac{1}{2}(\delta^k_p \delta^l_n
\eta_{qm}-\delta^l_p \delta^k_n \eta_{qm}-\delta^k_q \delta^l_n
\eta_{pm}+\delta^l_q \delta^k_n \eta_{pm}-\delta^k_p \delta^l_m
\eta_{qn}+\delta^l_p \delta^k_m \eta_{qn}+\delta^k_q \delta^l_m
\eta_{pn}-\delta^l_q \delta^k_m \eta_{pn}).$$ Then we can calculate the Casimir
operator $C_A$, defined by $C_A\, I_{PB} = f^K_{PM}f^K_{BM}$, where $C_A= 8$,
while the group unity in the adjoint representation is given by $I_{PB} =
\frac{1}{2} S_{pq,bc}$ in terms of $S$ defined in (\ref{unit}). Therefore, we
can expect that up to the one loop accuracy of pure gauge theory, the $\beta$
function is negative, and it is equal to $\beta(g^2_{\rm\scriptscriptstyle
Pl})=\frac{d g^2_{\rm\scriptscriptstyle Pl}}{d\ln\mu^2}= -\frac{11}{3}\,C_A
\frac{g^4_{\rm\scriptscriptstyle Pl}}{4\pi},$ pointing to the asymptotic
freedom.}, a study of quantum anomalies and a consequent motivated analysis of
vacuum structure.

This work is in part supported by the Russian Foundation for Basic Research,
grants 01-02-99315, 01-02-16585 and 00-15-96645.


\end{document}